\documentclass[a4paper]{article}
\usepackage[utf8]{inputenc}
\usepackage{amsmath,amssymb,amsfonts,amsthm}
\usepackage{mathrsfs}
\usepackage{graphicx,subfigure}
\usepackage{slashed}
\usepackage[nottoc]{tocbibind}
\usepackage{color} 
\usepackage{natbib}
\usepackage[hidelinks]{hyperref}

\usepackage{lscape}

\bibpunct{[}{]}{,}{n}{,}{,}

\newcommand{\nn}{\nonumber}
\newcommand{\hc}{\ensuremath{\mathrm{h.c.}}}
\newcommand{\U}[1]{\ensuremath{\mathrm{U(#1)}}}
\newcommand{\SU}[1]{\ensuremath{\mathrm{SU(#1)}}}
\newcommand{\lh}{\mathrm{L}}
\newcommand{\rh}{\mathrm{R}}

\newcommand{\diag}{\mathrm{diag}}

\newcommand{\mueg}{\mu\to e\gamma}
\newcommand{\mute}{\mu\to 3e}
\newcommand{\muec}{\mu\mbox{-}e}

\title{\bf Lepton flavor violation and scalar dark matter in a radiative model of neutrino masses}
\author{Michael Klasen, David R. Lamprea\\ \textit{\small Institut
für Theoretische Physik,}\\ \textit{\small Westfälische
Wilhelms-Universität Münster, }\\ \textit{\small Wilhelm-Klemm-Straße 9, D-48149
Münster, Germany}
\and Carlos E. Yaguna\\ \textit{\small Max-Planck-Institut fur Kernphysik,}\\ \textit{\small Saupfercheckweg 1, 69117 Heidelberg, Germany}}
\date{}
\begin{document}
\maketitle
\begin{abstract}
We consider a simple extension of the Standard Model that can account for the dark matter and explain the existence of neutrino masses. The model includes  a vector-like doublet of SU(2), a singlet fermion, and two scalar singlets, all of them odd under a new Z$_2$ symmetry. Neutrino masses are generated radiatively by one-loop processes involving the new fields, while the dark matter candidate is the lightest neutral  particle among them.  We focus specifically on the case where the dark matter particle is one of the scalars and its relic density is determined by its Yukawa interactions. The phenomenology of this setup, including neutrino masses, dark matter and lepton flavor violation, is analyzed in some detail. We find that the dark matter mass must be below $500$ GeV to satisfy the relic density constraint. Lepton flavor violating processes are shown to provide the most promising way to test this scenario. Future $\mu\to 3e$ and $\mu$-$e$ conversion experiments, in particular, have the potential to probe the entire viable parameter space of this model.
\end{abstract}

\section{Introduction}
\label{sec:intro}
Neutrino masses and dark matter provide compelling evidence for physics beyond the Standard Model (SM).  The gravitational effects of dark matter have been observed,  for instance,  in galaxies, clusters of galaxies, the large scale structure of the Universe, and in the cosmic microwave background radiation. Recently, the  WMAP \cite{Hinshaw:2012aka} and Planck \cite{Ade:2013zuv} collaborations have  determined the current dark matter density in the Universe to an unprecedented precision:  $\Omega_{DM}h^2=0.1186\pm 0.0031$. Oscillations experiments \cite{Fukuda:1998mi,Ahmad:2002jz,Araki:2004mb}, on the other hand,  have demonstrated that neutrinos have non-zero masses and have allowed us to measure the mixing and mass parameters in the neutrino sector \cite{Forero:2014bxa}. 

Within the SM, neither neutrino masses nor dark matter can be explained,  and  current data  does not tell us how the SM should be extended to account for them. A particularly appealing possibility is that these two problems are not independent, as usually assumed, but arise from the same type of  new physics. Moreover, this new physics may appear, as suggested by the WIMP paradigm of dark matter, at the TeV scale --the scale that is currently being probed by the LHC. In \cite{Restrepo:2013aga}, many models of this type, featuring neutrino masses at one-loop,  were found and classified.  They all contain a small number of additional fields and a new discrete symmetry to stabilize the dark matter particle. In this paper, we study one of these models, denoted as T1-3 in \cite{Restrepo:2013aga}. 

In this model, which had been previously considered in \cite{Fraser:2014yha,Restrepo:2015ura}, the SM is extended with a vector-like SU(2) doublet, a singlet fermion, and two singlet scalars. We focus specifically on the case where the dark matter particle is a scalar and its relic density is determined by the Yukawa interactions (rather than the scalar ones), a possibility that to our knowledge has not been studied before. As we will show, in this setup strong correlations arise between dark matter, neutrino masses and lepton flavor violating (LFV) processes. In our analysis, we will  investigate these correlations in some detail, partly relying on a scan over the parameter space of this model.  First, we  use a large sample of viable models to determine and study the regions that  are consistent with the constraints from neutrino masses, $\mueg$, and dark matter. Then, we review the  predictions for other LFV processes and  the prospects for detection in future experiments. We find, in particular, that future searches for $\mute$ and $\muec$ conversion in nuclei have the potential to probe the entire parameter space of this model. We also  discuss alternative ways to test this scenario, including collider searches as well as direct and indirect dark matter searches.

The rest of the paper is organized as follows. In the next section we describe the model and introduce our notation. Then, in Sec.\ \ref{sec:pheno} we analyze semi-quantitatively the most relevant phenomenological aspects of this model, including neutrino masses, LFV processes, and dark matter. Our main results, based on a scan over the parameter space of this model, are presented in Sec.\ \ref{sec:results}. First we examine the viable parameter space and then we demonstrate that the rates for several LFV processes are significant,  offering the potential to probe this model thoroughly in the near future. In Sec.\ \ref{sec:dis} we briefly discuss the  possibility to test this scenario using collider or dark matter searches.  Finally, we summarize and draw our conclusions in Sec.\ \ref{sec:conc}.

\section{Description of the model}
\label{sec:model}
The model we consider belongs, in the generic classification of \cite{Bonnet:2012kz}, to the  T1-iii  models of one-loop neutrino masses. It corresponds, in particular, to the model T1-3-A with $\alpha=0$ (where $\alpha$ is related to the hypercharge), as defined  in \cite{Restrepo:2013aga}. In this model the SM particle content is extended with a vector-like fermion  (or two chiral fermion) doublet under SU(2),  $D,D^\prime$, one left-handed singlet fermion, $S$, and two real scalar singlets $\phi_i$ ($i=1,2$). To guarantee the stability of the dark matter candidate and to prevent tree-level neutrino masses, the SM gauge group  is extended with a Z$_2$ discrete symmetry, under which all the new fields are odd, while the SM fields are even. Thus, the charges of the new particles under the $\SU{2}\otimes\U{1}\otimes$ Z$_2$ symmetry are given by

\begin{gather}
  \label{eq:charges-doublets}
  D =
  \left(
    \begin{array}{c}
      \psi\\
      E
    \end{array}
  \right)
  \sim (2, -\frac{1}{2}, -), \quad
  D^\prime =
  \left(
    \begin{array}{c}
      -E^\prime\\
      \psi^\prime
    \end{array}
  \right)
  \sim (2, \frac{1}{2}, -),\nn\\
\end{gather}
for the fermion doublets and
\begin{equation}
  \label{eq:charges-singlets}
  S \sim (1, 0, -), \quad \phi_i \sim (1, 0, -)
\end{equation}
for the fermion and scalar singlets, respectively. The SM electroweak sector under the same symmetry transforms instead as

\begin{gather}
  \label{eq:smcharges-doublets}
  H =
  \left(
    \begin{array}{c}
      \phi^+\\
      \phi^-
    \end{array}
  \right)
  \leadsto
  \left(
    \begin{array}{c}
      0\\
      \frac{1}{\sqrt{2}} (v + h)
    \end{array}
  \right)
  \sim (2, \frac{1}{2}, +), \quad
  L_i =
  \left(
    \begin{array}{c}
      \nu_{i\lh}\\
      e_{i\lh}
    \end{array}
  \right)
  \sim (2, -\frac{1}{2}, +),
\end{gather}
for the doublets ($i=1...3$) and
\begin{equation}e_{i\rh} \sim (1, 1, +)\end{equation}
for the SU(2) singlets. The  most general Lagrangian consistent with the symmetry and the particle content of our model can be written as
\begin{align}
  \label{eq:lagrangian}
  \mathscr{L}  = & \mathscr{L}_{SM}- V(H,\phi_i)\nonumber\\
  &-\alpha_{ij} \overline{D^\prime} L_j^c \phi_i - \mu \overline{D^c} D^\prime - \beta_1 \overline{D^c} HS -
  \beta_2 \overline{D^\prime} \tilde H S^c  - \frac{1}{2}m_S \overline{S^c} S + \hc,
\end{align}
where $\mathscr{L}_{SM}$ is the SM Lagrangian, $V(H,\phi_i)$ is the scalar potential, and  $\tilde H = i\sigma_2 H^\ast$. This Lagrangian includes a Dirac mass term for the vector-like doublet fermion, a Majorana mass term for the singlet fermion,  Yukawa terms between the new singlet and doublet fermions,  and a new Yukawa interaction for  the SM lepton doublets.

It is precisely this new interaction that allows to obtain non-zero neutrino masses. At tree-level, neutrinos remain massless  because the scalar fields $\phi_i$ do not acquire a vev  -- the Z$_2$ symmetry must remain exact to explain the dark matter. It is only at one-loop that neutrinos will get a Majorana mass.  In fact, it is not difficult to verify that this Lagrangian generically violates lepton number. Being real scalar fields, $\phi_i$ should have zero lepton number and likewise the singlet Majorana fermion $S$. But then the terms with coefficients $\alpha_{ij}$, $\beta_1$ and $\beta_2$ cannot simultaneously conserve  lepton number. Lepton number would still be a good symmetry  in the limits $\alpha_{ij}=0$  or $\beta_1=\beta_2=0$. Neutrino masses, which are lepton-number violating, must therefore vanish in such cases.

Once expanded, the Lagrangian includes the following mass terms
\begin{align}
  \label{eq:lagrangian-mass}
  \mathcal{L}_m & =
  -\mu \overline{\psi^c} \psi^\prime - \mu \overline{E^c} E' - \frac{v \beta_1}{\sqrt{2}}\overline{\psi^c} S
  \nonumber\\
  & - \frac{v \beta_2}{\sqrt{2}} \overline{\psi^\prime} S^c - \frac{1}{2} m_S \overline{S^c} S - \frac{1}{2} m_{\phi_i} \phi_i^2 + \hc
\end{align}
and the following couplings with the SM particles
\begin{align}
  \label{eq:l-couplings}
  \mathcal{L}_Y & =
  \alpha_{ij} \overline{\psi^\prime} \nu_{j\lh}^c \phi_i + \alpha_{ij}  \overline{E^\prime} e_{jL}^c \phi_i + \frac{\beta_1}{\sqrt{2}} \overline{\psi^c} S h + \frac{\beta_2}{\sqrt{2}} \overline{\psi^\prime} S^c h + \hc
\end{align}

Defining $\Psi := (\psi, \psi^\prime, S)$, the  Majorana mass matrix for the odd neutral fermions can be written, from Eq.\ \eqref{eq:lagrangian-mass}, as
\begin{equation}
  \label{eq:mass-matrix}
  m_\Psi = 
  \left(
    \begin{matrix}
      0 & \mu & \frac{v\beta_1}{\sqrt{2}}\\
      \mu & 0 & \frac{v\beta_2}{\sqrt{2}}\\
      \frac{v\beta_1}{\sqrt{2}} & \frac{v\beta_2}{\sqrt{2}} &  m_S
    \end{matrix}
  \right).
\end{equation}
This matrix can be diagonalized via the transformation
\begin{equation}
  \Psi_i \equiv \xi_{ij} \chi_j,
\end{equation}
where $\xi$ is the mixing matrix and $\chi_i$ are the physical mass eigenstates with masses
$m_i$ such that $m_i < m_j$ for $i < j$.

In addition to these three Majorana fermions, the spectrum contains also a charged Dirac fermion with mass $\mu\equiv m_E$  -- see Eq.\ \eqref{eq:lagrangian-mass} --  constrained by collider searches \cite{Agashe:2014kda} to be larger than about $103.5$ GeV, and two neutral scalars with masses $m_{\phi_{1,2}}$.

\section{Phenomenology}
\label{sec:pheno}
In this section,  the main phenomenological features of this model are discussed, in particular regarding neutrino masses, dark matter, and lepton flavor violating processes.

\subsection{Neutrino masses}
\label{sec:neutrinomasses}

\begin{figure}
  \centering
  \resizebox{5cm}{!}{\includegraphics{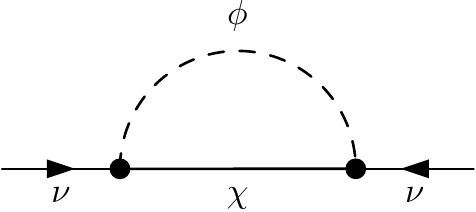}}
  \resizebox{5cm}{!}{\includegraphics{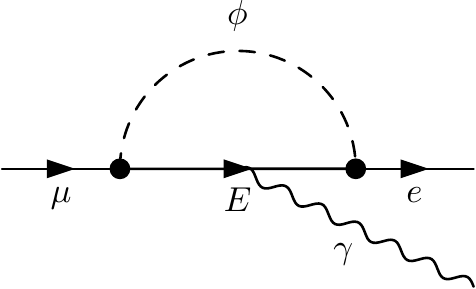}}
  \caption{Left: The diagram responsible for neutrino mass generation. Right: One of the diagrams contributing to $\mueg$.}
  \label{fig:diagrams}
\end{figure}
Radiative mechanisms are a very attractive way of explaining neutrino masses and have been implemented within many different scenarios  -- see e.g. \cite{Zee:1980ai,Babu:1988ki,Krauss:2002px,Ma:1998dn,Ma:2006km,Farzan:2009ji,Dev:2012sg,Farzan:2012ev,Law:2013saa,Hirsch:2013ola,Restrepo:2015ura,Longas:2015sxk}. In the model we consider, neutrino masses are obtained through a radiative  mechanism at one-loop. The relevant diagram, shown in the left panel of Fig.~\ref{fig:diagrams}, contains the neutral odd particles (scalars and fermions) running in the loop. Since one of these neutral particles will be the dark matter candidate, some connection between dark matter and neutrino masses is necessarily present. The evaluation of the diagram yields, for a given neutral fermion ($k$) and a given singlet scalar ($r$),  the following  contribution to the neutrino mass matrix
\begin{multline}
  \label{eq:matrix-element}
  M_{ij}^{k,r} = \frac{1}{16\pi^2} \Gamma_{jk}^{r} \Gamma_{ik}^{r} m_k\\
  \times \left( (1+\Delta) - \frac{m_k^2}{m_k^2 - m_r^2}\log\left(\frac{m_k^2}{\mu_R^2}\right) + \frac{m_r^2}{m_k^2 - m_r^2}\log\left(\frac{m_r^2}{\mu_R^2}\right) \right),
\end{multline}
where  $\Gamma_{lm}^r = \alpha_{rl}\xi_{2m}$ is the coupling between $\phi_r$, $\chi_m$, and $\nu_\lh^c$. In this equation, $\Delta \propto \frac{1}{\epsilon}$ ($\epsilon=4-d$) whereas $\mu_R$ is an arbitrary renormalization constant. The neutrino mass matrix element $m^\nu_{ij}$  can therefore be computed as a sum over all the fermions and scalars that run in the loop:
\begin{equation}
  \label{eq:neutrino-mass-matrix}
  m^\nu_{ij} = \sum_{k,r} M_{ij}^{k,r}
  = \sum_{k,r} \frac{1}{16\pi^2} m_k
  \alpha_{ri} \alpha_{rj} \xi_{2k} \xi_{2k}
  \left( \frac{m_k^2}{m_k^2 - m_r^2} \log\left(\frac{m_k^2}{m_r^2}\right) \right),
\end{equation}
where we have used the fact that
\begin{equation}
  \sum_k \xi_{2k} m_k \xi^T_{k2} = (m_\Psi^T)_{22} = 0.
\end{equation}

Notice that, as expected, the final result is finite and thus independent of the unphysical renormalization factor $\mu_R$. As anticipated, the neutrino mass matrix vanishes in the limit $\alpha_{ij}=0$, where lepton number is conserved. To see that it also vanishes in the limit $\beta_1=\beta_2=0$, we can diagonalize Eq.\ \eqref{eq:mass-matrix} analytically at leading order in $\beta_1,\beta_2\ll 1$ to obtain

\begin{equation}
  \xi =
  \begin{pmatrix}
    \frac{i}{\sqrt{2}} & \frac{1}{\sqrt{2}} & \frac{2m_S\beta_1 + 4\mu\beta_2}{m_S^2 - 4\mu^2} v\\
    -\frac{i}{\sqrt{2}} & \frac{1}{\sqrt{2}} & \frac{2m_S\beta_2 + 4\mu\beta_1}{m_S^2 - 4\mu^2} v\\
    i\frac{\beta_2 - \beta_1}{m_S + 2\mu} \frac{v}{\sqrt{2}} & \frac{\beta_1+\beta_2}{2\mu - m_S} \frac{v}{\sqrt{2}} & 1
  \end{pmatrix}
   + \mathcal{O}(\beta_1^2,\beta_2^2,\beta_1\beta_2).
\end{equation}
Hence,  the neutrino masses in Eq.\ \eqref{eq:neutrino-mass-matrix} can be expressed as
\begin{equation}
\label{eq:mnu}
  m^\nu_{ij} =
  \sum_{r} \frac{1}{16\pi^2} m_S
  \alpha_{ri}\alpha_{rj} \left(\frac{2m_S\beta_2 + 4\mu\beta_1}{m_S^2 - 4\mu^2}\right)^2 v^2
  \left( \frac{m_S^2}{m_S^2 - m_r^2} \log\left(\frac{m_S^2}{m_r^2}\right) \right),
\end{equation}
which indeed goes to zero in the limit $\beta_{1,2}\to 0$. Since the experimentally determined neutrino masses are tiny, either $\alpha_{ij}$ or $\beta_{1,2}$ must be  suppressed. To estimate the typical values of the parameters that give rise to neutrino masses consistent with the data, we can evaluate Eq.\ \eqref{eq:mnu} for masses around the TeV scale to obtain
\begin{equation}
 m^\nu\sim 0.1~\mathrm{eV}\times\left(\frac{m_S}{1~\mathrm{TeV}}\right) \left(\frac{\alpha_{ij} 
 \beta_{1,2}}{10^{-5}}\right)^2.
\end{equation}
Notice that neutrino masses cannot distinguish whether it is the $\alpha_{ij}$ or the $\beta_{1,2}$ that are suppressed but, as we will see later, the dark matter constraint only allows the latter option. 

Since $\alpha_{ij}$ and $\beta_{1,2}$ enter both quadratically in $m^\nu$, the required coupling suppression is not as large as in the well-known scotogenic model \cite{Ma:2006km}, where $\lambda_{5}$ can be of order $10^{-10}$ \cite{Vicente:2014wga}. In our model the coupling suppression is only of order $10^{-5}$.  

It is easy to verify that, in our scenario, the neutrino mass matrix has only two non-zero eigenvalues. This result is a direct consequence of the two real scalar fields that interact with the lepton doublets. To obtain three non-zero neutrino masses,  one more scalar field would be needed.  But, since current neutrino data is consistent with one massless neutrino, we will stick to this minimal framework, as outlined in the previous section. The neutrino spectrum, therefore, is necessarily hierarchical and we assume it henceforth to be of normal type (NH).

The resulting neutrino mass matrix, Eq.\ (\ref{eq:neutrino-mass-matrix}), can be written in a way reminiscent of the seesaw mechanism as
\begin{equation}
  m_{ij}^\nu = \alpha^{\rm T}_{ir} F_r \alpha_{rj},
\end{equation}
with
\begin{equation}
  F_r
  = \sum_k \frac{1}{16\pi^2} m_k
  \xi_{2k} \xi_{2k}
  \left( \frac{m_k^2}{m_k^2 - m_r^2} \log\left(\frac{m_k^2}{m_r^2}\right) \right).
\end{equation}
This form of the neutrino mass matrix allows us to use a slightly modified version of the Casas-Ibarra parametrization~\cite{Casas:2001sr}. Explicitly, we have that
\begin{equation}
  \label{eq:matrices}
  \alpha = i U^* \sqrt{m^\nu} R \sqrt{F^{-1}},
\end{equation}
where $m^\nu=\mathrm{diag}(0, m_2^\nu, m_3^\nu)$ are the light neutrino masses, $U$ is the PMNS neutrino mixing matrix,  $F=\diag(F_1, F_2)$, and $R$ is a $3\times 2$ matrix such that $R R^T=1$. This matrix depends on a single parameter, $\theta$, as \cite{Ibarra:2003up}
\begin{equation}
  R = \begin{pmatrix}
  0 & 0\\
\cos\theta & \pm \sin\theta\\
-\sin\theta & \pm \cos\theta
\end{pmatrix}.
\end{equation}

Equation \eqref{eq:matrices} provides, for a given set of  scalar and fermion mass parameters, the most general form of the $\alpha_{ij}$ couplings that is consistent with the observed neutrino data, which are used as input parameters.  In other words, the constraints on neutrino masses and mixing angles are automatically incorporated into the structure of $\alpha$ as given by Eq.\ \eqref{eq:matrices}, simplifying enormously the analysis of the viable parameter space.

\subsection{Lepton flavor violating processes}
The existence of non-zero neutrino masses and mixing angles also imply the violation of the lepton flavor. In consequence,  processes involving the charged leptons and where the lepton flavor is not conserved  -- lepton flavor violating (LFV) processes --  such as $\mu\to e\gamma$ and $\mu\to 3e$ should occur at some level in this model.  The relevant diagrams  appear first at one-loop, as illustrated in the right panel of Fig.\ \ref{fig:diagrams} for $\mu\to e \gamma$. Notice that the fermion running in the loop is now the Dirac charged fermion rather than the neutral ones.

In this model, the $\mu\to e\gamma$ branching ratio  can be evaluated to be
\begin{equation}
  {\rm BR}(\mu\to e\gamma) = \frac{3\alpha_{\rm em}}{64\pi G_{\rm F}^2 m_E^4} \left|\sum_r\alpha_{r,1}\alpha_{r,2} G(m_{\phi_r}^2/m_E^2)\right|^2,
\end{equation}
where
\begin{equation}
  G(x) = \frac{2 - 3x - 6x^2 + x^3 + 6x\log x}{6(1-x)^4}.
\end{equation}
An analogous expression holds for the related processes $\tau\to \mu\gamma$ and $\tau\to e \gamma$. For other processes, the analytical expressions are more complicated and are therefore better studied numerically (with the help of FlavorKit \cite{Porod:2014xia}), as we will do in the next section. 
\begin{table}[tb!]
\centering
\begin{tabular}{|c|c|c|}
\hline
LFV Process & Present Bound & Future Sensitivity  \\
\hline
    $\mu \rightarrow  e \gamma$ & $5.7\times 10^{-13}$~\cite{Adam:2013mnn}  & $6\times 10^{-14}$~\cite{Baldini:2013ke} \\
    $\tau \to e \gamma$ & $3.3 \times 10^{-8}$~\cite{Aubert:2009ag}& $ \sim3\times10^{-9}$~\cite{Aushev:2010bq}\\
    $\tau \to \mu \gamma$ & $4.4 \times 10^{-8}$~\cite{Aubert:2009ag}& $ \sim3\times10^{-9}$~\cite{Aushev:2010bq} \\
    $\mu \rightarrow e e e$ &  $1.0 \times 10^{-12}$~\cite{Bellgardt:1987du} &  $\sim10^{-16}$~\cite{Blondel:2013ia} \\
    $\tau \rightarrow \mu \mu \mu$ & $2.1\times10^{-8}$~\cite{Hayasaka:2010np} & $\sim 10^{-9}$~\cite{Aushev:2010bq} \\
    $\tau^- \rightarrow e^- \mu^+ \mu^-$ &  $2.7\times10^{-8}$~\cite{Hayasaka:2010np} & $\sim 10^{-9}$~\cite{Aushev:2010bq} \\
    $\tau^- \rightarrow \mu^- e^+ e^-$ &  $1.8\times10^{-8}$~\cite{Hayasaka:2010np} & $\sim 10^{-9}$~\cite{Aushev:2010bq} \\
    $\tau \rightarrow e e e$ & $2.7\times10^{-8}$~\cite{Hayasaka:2010np} &  $\sim 10^{-9}$~\cite{Aushev:2010bq} \\
    $\mu^-, \mathrm{Ti} \rightarrow e^-, \mathrm{Ti}$ &  $4.3\times 10^{-12}$~\cite{Dohmen:1993mp} & $\sim10^{-18}$~\cite{PRIME,Sato:2008zzm} \\
    $\mu^-, \mathrm{Au} \rightarrow e^-, \mathrm{Au}$ & $7\times 10^{-13}$~\cite{Bertl:2006up} & \\
    $\mu^-, \mathrm{Al} \rightarrow e^-, \mathrm{Al}$ &  & $10^{-15}-10^{-18}$~\cite{Litchfield:2014qea}\\
    $\mu^-, \mathrm{SiC} \rightarrow e^-, \mathrm{SiC}$ &  & $10^{-14}$~\cite{Natori:2014yba} \\
\hline
\end{tabular}
\caption{Current experimental bounds and future sensitivities for the most important LFV observables. \label{table}}
\end{table}

Typically, $\mueg$  provides the  most stringent constraint among  LFV processes, so many works have focused on this process. But, as emphasized in \cite{Vicente:2014wga}  this situation might drastically change in the near future thanks to the significant improvements that will likely be achieved for other LFV processes, as shown in Tab.\ \ref{table}. Particularly relevant will be the expected limits on $\mu\to 3e$ and $\mu$-$e$ conversion in nuclei, which may improve, respectively, by up to four and six orders of magnitude \cite{Mihara:2013zna}. It is important, therefore, not to limit the discussion on LFV processes to $\mu\to e\gamma$ but to consider also these other promising processes when analyzing future detection prospects. 

A crucial difference between neutrino masses and LFV processes, both of which appear at one-loop in this model, is that the former violate lepton number, while the latter do not. As a result, LFV processes depend only on the $\alpha_{ij}$ couplings but not on $\beta_{1,2}$ or $m_S$, as shown explicitly  above for $\mu\to e\gamma$.

To obtain observable rates for LFV processes, two conditions must generically be satisfied: the particles in the loop must no be that heavy and the relevant couplings,  $\alpha_{ij}$ in this case, should be of order $1$-$0.1$. As we saw in the previous subsection, this second requirement is compatible with neutrino masses as long as the beta parameters are tiny, $\beta_{1,2}\ll 1$. Next, we show that, under certain circumstances, these requirements are actually enforced by the dark matter constraint.

\subsection{Dark matter phenomenology}
The dark matter candidate in this model is the lightest neutral odd  (under the $Z_2$) particle in the spectrum, which can be one of the three fermions, or one of the two scalars. Since the fermion and scalar masses are free parameters, both of these possibilities can be realized in this  model. When the dark matter is a fermion, the resulting phenomenology is very similar to that of the so-called singlet-doublet fermion model \cite{ArkaniHamed:2005yv, Mahbubani:2005pt,D'Eramo:2007ga}, which has been extensively studied in the recent literature  -- see e.g. \cite{Enberg:2007rp,Cohen:2011ec,Cheung:2013dua,Calibbi:2015nha,Yaguna:2015mva}. It has been shown, in particular, that by adjusting the mixing between the singlet and the doublet one can obtain the correct relic density, via freeze-out in the early Universe, for dark matter masses below $1~\mathrm{TeV}$ or so. The additional scalars present in this model may sligthly modify this picture due to coannihilation effects, as recently demonstrated in \cite{Restrepo:2015ura}.

\begin{figure}[tb]
  \centering
  \includegraphics[width=5cm]{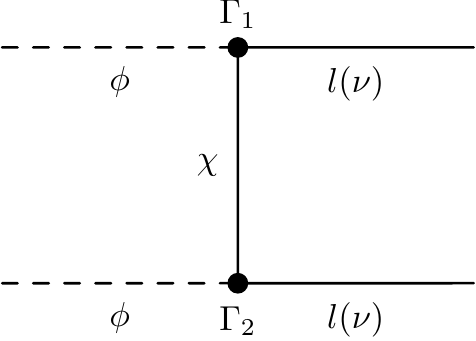}
  \caption{General diagram contributing to the relic density computation.}
  \label{fig:relic-comparison}
\end{figure}

When the dark matter is the scalar, two different scenarios can be distinguished depending on its dominant interactions. Scalars, in fact, not only have the Yukawa interactions explicitly shown in Eq.\ (\ref{eq:l-couplings}) but also scalar interactions with the SM Higgs boson, implicitly included in  $V(H, \phi_{i})$.  These scalar interactions give rise to the so-called singlet scalar or Higgs-portal model \cite{Silveira:1985rk,McDonald:1993ex,Burgess:2000yq}, where dark matter annihilations are mediated by the Higgs boson. These models are quite predictive and have been the subject of many previous analyses \cite{Davoudiasl:2004be,Barger:2007im,Dick:2008ah,Yaguna:2008hd,Goudelis:2009zz,Profumo:2010kp,Yaguna:2010hn,Yaguna:2011qn,Mambrini:2011ik,Cline:2013gha}. In this paper we assume instead that the scalar dark matter  interacts dominantly via the Yukawa interactions in Eq.\ \ref{eq:l-couplings}. Thus, it will annihilate into leptons via $t$-channel fermion-mediated diagrams with a cross section proportional to $\alpha_{ij}^4$  -- see Fig.\ \ref{fig:relic-comparison}. Consequently, non-trivial correlations between neutrino masses, LFV processes and dark matter are expected. In fact, we can already state that to obtain a relic density in agreement with the observations, $\alpha_{ij}$ should be of order one and the mediators cannot be that heavy, which are essentially the same conditions that ensure observable rates for LFV processes, as we saw in the previous subsection. Throughout the rest of the paper, we will be working on this specific framework where the dark matter particle is a scalar, denoted by $\phi_1$, that interacts dominantly via the Yukawa terms with the SM lepton doublets. Notice that as a result, the dark matter turns out to be \emph{leptophilic}, with important implications for  the direct detection prospects, as will be discussed in Sec.\ \ref{sec:dis}.

In this setup, dark matter annihilates into charged leptons with a cross section given, in the non-relativistic limit, by \cite{Giacchino:2013bta}
\begin{equation}
  \sigma v (\phi_1\phi_1\to \ell^+\ell^-)= \frac{\alpha_{1\ell}^4 v^4}{60\pi}\frac{m_{\phi_1}^2}{m_{\phi_1}^4+m_E^4}.
\end{equation}
This cross section gets suppressed as the dark matter mass or the charged fermion mass increases. Notice also that it has a strong velocity dependence ($\propto v^4$). During freeze-out, this velocity suppression  is not that important but today, when $v\sim 10^{-3}$ in our Galaxy, it pretty much prevents the annihilation of  dark matter particles into two charged leptons, significantly affecting the dark matter indirect detection signatures in this model.

The annihilation into neutrinos proceeds through a similar diagram, but it turns out to be negligible due to the fact that the exchanged fermion is a Majorana rather than a Dirac particle. Thus, only charged leptons contribute to the total annihilation rate. In our numerical analysis we will see that, due to the constraints from $\mu\to e\gamma$, it is the $\tau^+\tau^-$ final state that actually dominates the cross section.

To obtain an accurate prediction of the dark matter relic density, we implemented this model into DarkSUSY \cite{Gondolo:2004sc} (micrOmegas \cite{Belanger:2013oya}  gives numerical errors in the evaluation of the annihilation cross section). Specifically, we use the $\sigma v$ above  to compute the invariant annihilation rate as defined in the DarkSUSY manual, and use the function {\tt dsrdens()} to obtain the relic density. We also verified that the computations done in this way  match the  approximate analytical results known in the literature.

\section{Numerical results}
\label{sec:results}

In this model  non-trivial correlations between dark matter, neutrino masses and lepton flavor violating processes are expected. The reason is that, on the one hand, the dark matter constraint requires sizable $\alpha_{ij}$ couplings between the dark matter and the SM leptons. On the other hand, these couplings cannot be flavor diagonal because they determine the structure of the neutrino mass matrix. Consequently, these couplings induce significant rates for LFV processes.  In this section, we use a scan over the parameter space of this model to numerically  study these correlations. To that end, we first obtain a large sample of viable models that we use to analyze the regions in the parameter space of the model that are consistent with current bounds. Then, we analyze the prediction for  LFV processes and demonstrate that future LFV experiments have the potential to probe most of the viable models.

\subsection{The viable parameter space}
\begin{figure}[tb]
  \centering
  \includegraphics[scale=0.45]{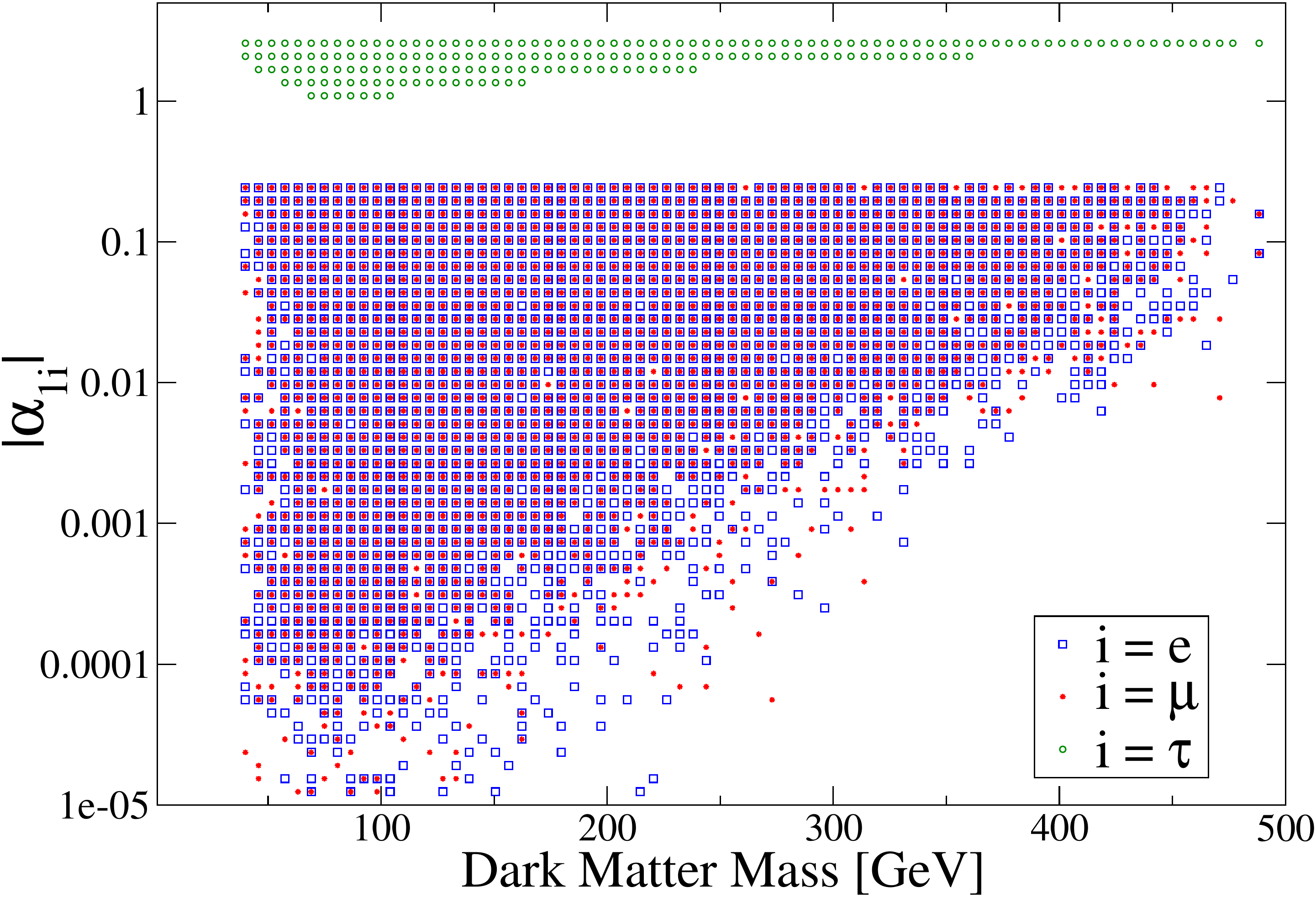}
  \label{fig:alpha}
  \caption{The viable models projected onto the plane ($m_\phi$,$\alpha_{i1}$). The $\alpha_{i1}$ couplings determine the dark matter annihilation rate and must therefore be non-negligible.}.
\end{figure}

The relevant free parameters of this model  are just seven: $m_{\phi_{1,2}}$, $\mu$, $m_S$, $\beta_{1,2}$ and $\theta$.  Our sample of viable models is obtained after scanning over the allowed range of these parameters and imposing the following constraints: the dark matter relic density ($\Omega_{DM}h^2\sim 0.12$), the $\mu \to e \gamma$ upper limit  -- BR($\mu\to e\gamma)<5.7\times 10^{-13}$ -- , and perturbativity (somewhat arbitrarily, we require all dimensionless couplings to be smaller than three). Regarding the dark matter relic density, we assume that it is obtained via thermal freeze-out in the early Universe and that coannihilation effects  play no role in its determination, as is generically the case. To enforce this latter condition, we require all other masses to be larger than $1.2m_{\phi_1}$. The upper bound on the mass parameters is taken to be $10$ TeV.  By construction, all our models are consistent with neutrino data, for we use as input the experimental data on neutrino masses and mixing angles according to Eq.\ \eqref{eq:matrices}. In our analysis, we made use of SARAH and the BSM Tool Box scripts~\cite{Staub:2011dp}, which facilitate the whole process.

\begin{figure}[tb]
  \begin{tabular}{cc}
  \includegraphics[scale=0.23]{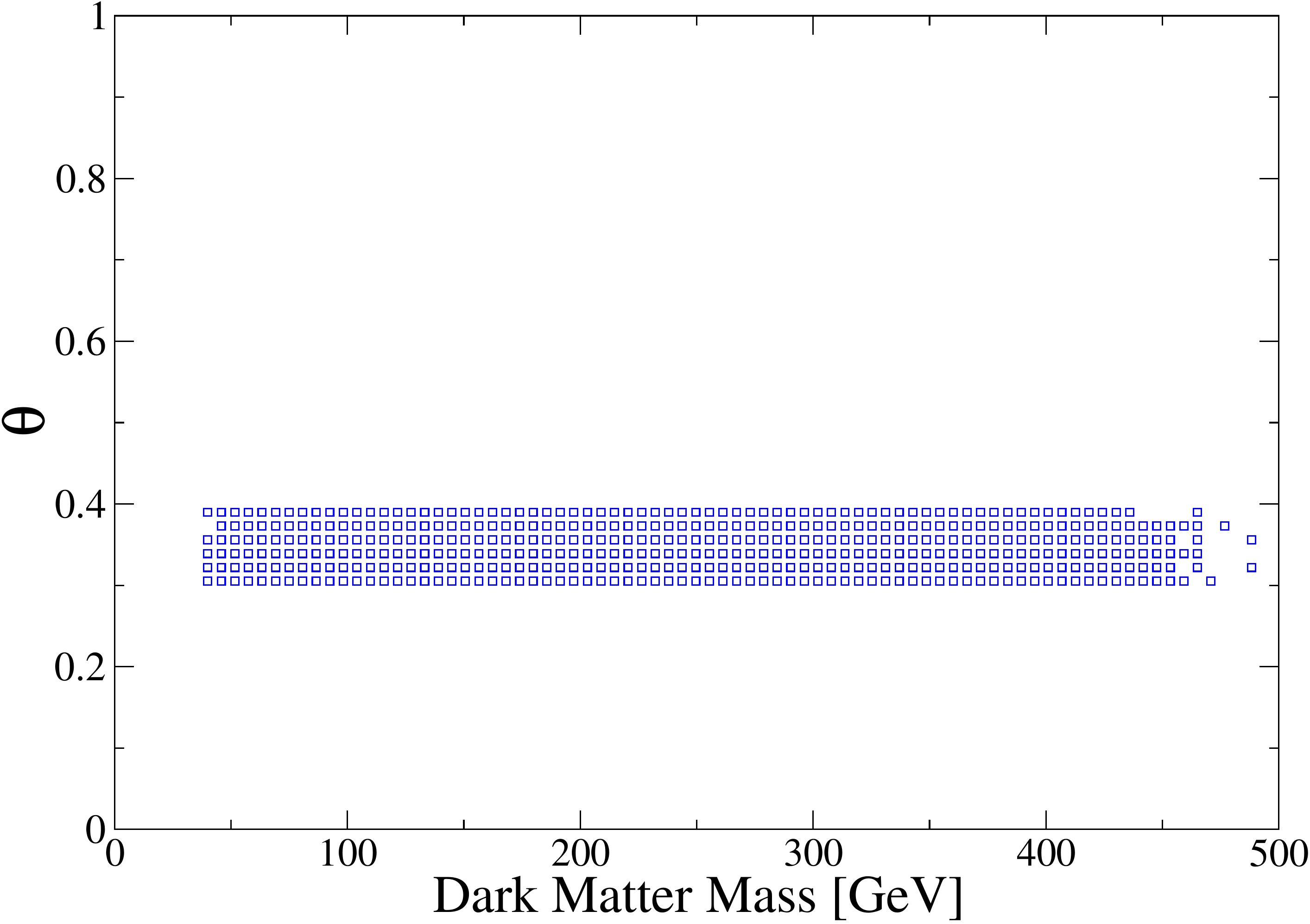} & \includegraphics[scale=0.23]{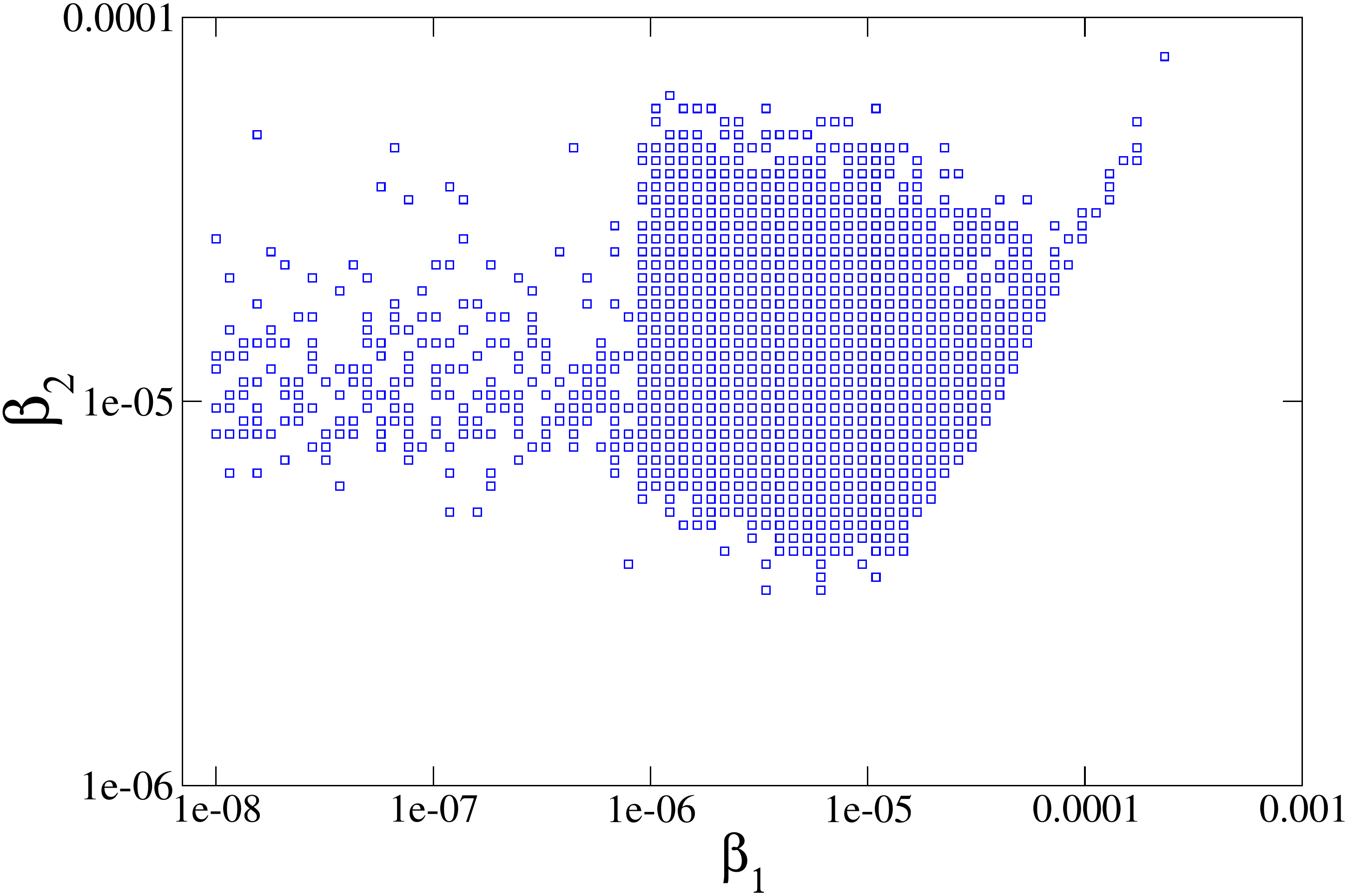}
  \end{tabular}
  \label{fig:theta}
  \caption{Left: The viable models projected onto the plane ($m_\phi$,$\theta$). Notice that the experimental constraints essentially select a single value of $\theta$. Right: The viable models projected onto the plane ($\beta_1$, $\beta_2$). Neutrino masses require $\beta_{1,2}$ to be small. }
\end{figure}

Let us begin by examining the viable regions of the parameter space. Figure 3 displays our set of viable models in the plane ($m_\phi$,$\alpha_{i1}$), with $i=e,\mu,\tau$. Notice that, as a result of the strong bounds from $\mueg$,  $\alpha_{\tau 1}\gg\alpha_{e 1},\alpha_{\mu 1}$. Thus, dark matter annihilates dominantly into the $\tau^+\tau^-$ final state. In fact, $\alpha_{\tau 1}$ increases with the dark matter mass, going from about one for $m_\phi\sim 50~\mathrm{GeV}$ to three  -- the perturbativity limit we imposed --  for $m_\phi\sim 500~\mathrm{GeV}$. Consequently, the dark matter particle in this scenario must be light, lying below $500~\mathrm{GeV}$.

The condition $\alpha_{1\tau}\gg\alpha_{1e },\alpha_{1\mu}$ is not satisfied for generic values of the $\alpha_{ij}$ couplings. According to Eq.\ \eqref{eq:matrices}, we have that
\begin{equation}
\frac{\alpha_{1\tau}}{\alpha_{1e}} =
\frac{\sqrt{m_2^\nu}\cos(\theta)(-c_{23}s_{12}s_{13} - c_{12}s_{23}) - \sqrt{m_3^\nu}\sin(\theta)c_{13}c_{23}}{\sqrt{m_2^\nu}\cos(\theta) c_{13}s_{12} - \sqrt{m_3^\nu}\sin(\theta)s_{13}},
\end{equation}
and
\begin{equation}
\frac{\alpha_{1\tau}}{\alpha_{1\mu}} =
\frac{\sqrt{m_2^\nu}\cos(\theta)(-c_{23}s_{12}s_{13} - c_{12}s_{23}) - \sqrt{m_3^\nu}\sin(\theta)c_{13}c_{23}}{\sqrt{m_2^\nu} (c_{12} c_{23} - s_{12} s_{13} s_{23}) \cos(\theta) - c_{13} \sqrt{m_3^\nu} s_{23}\sin(\theta)}.
\end{equation}
If we now require $\alpha_{1\tau}\gg\alpha_{1e },\alpha_{1\mu}$ we get that, for typical values of the neutrino parameters, $\theta\sim 0.35$. The left panel of Fig.\ 4 shows the value of $\theta$ for our set of viable models. We see that, indeed, $\theta$ varies only within a narrow range around $0.35$. Thus, the dark matter constraint and the limits on LFV processes select a rather specific value for $\theta$.

As we saw in the previous section, the parameters $\beta_1$ and $\beta_2$ must be tiny so that neutrino masses are sufficiently suppressed.  The right panel of Fig.\ 4 displays the viable models in the plane ($\beta_1$, $\beta_2$) and confirms that this is really the case. These parameters are never larger than about $10^{-4}$. $\beta_1$ could be much smaller than that, whereas $\beta_{2}$ varies only between $10^{-4}$ and about $10^{-6}$. Notice that the dark matter constraint plays also a role in this case, as it enforces a sizable value of the $\alpha_{ij}$ couplings.

\begin{figure}[tb]
  \begin{tabular}{cc}
  \includegraphics[scale=0.23]{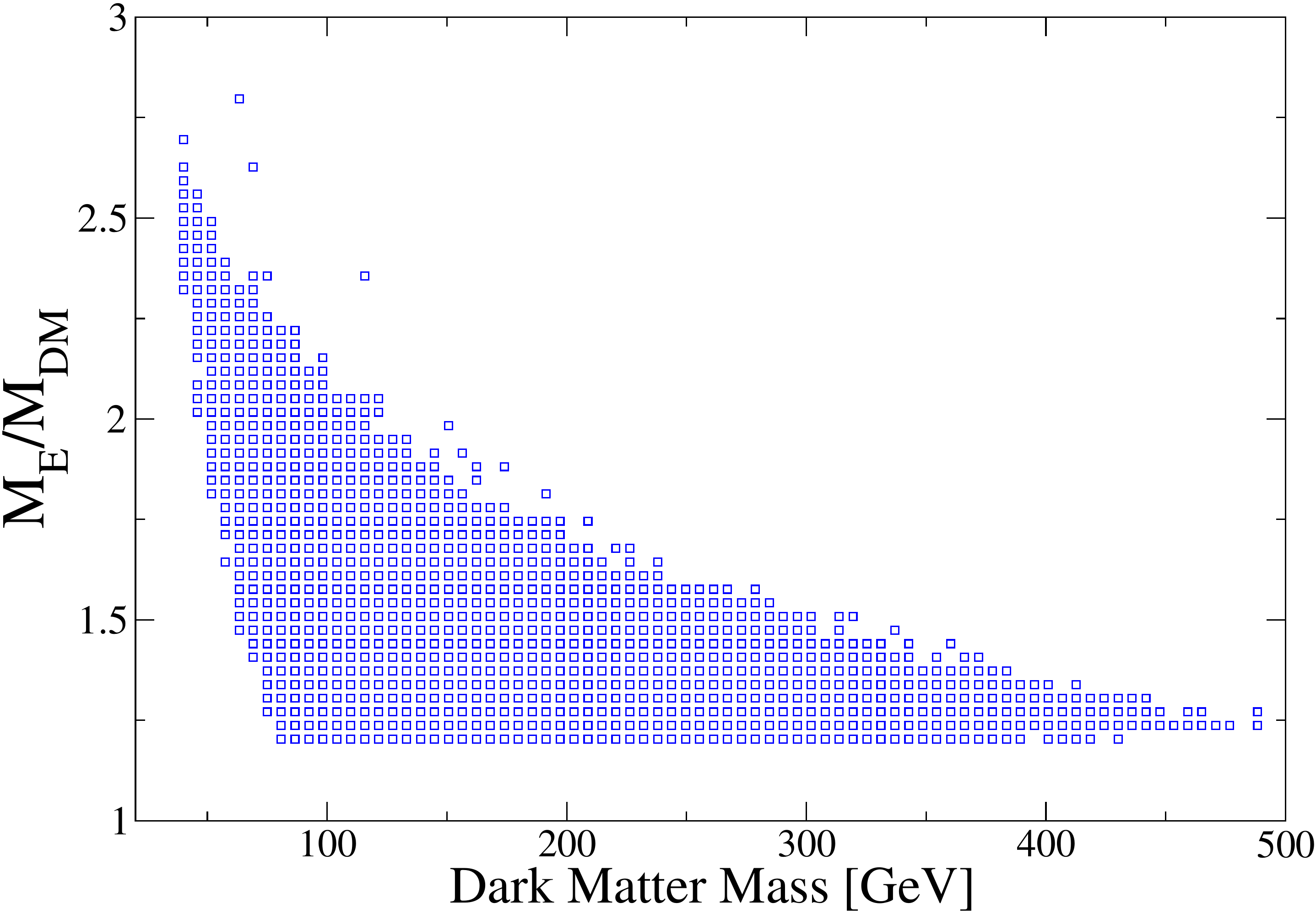} & \includegraphics[scale=0.23]{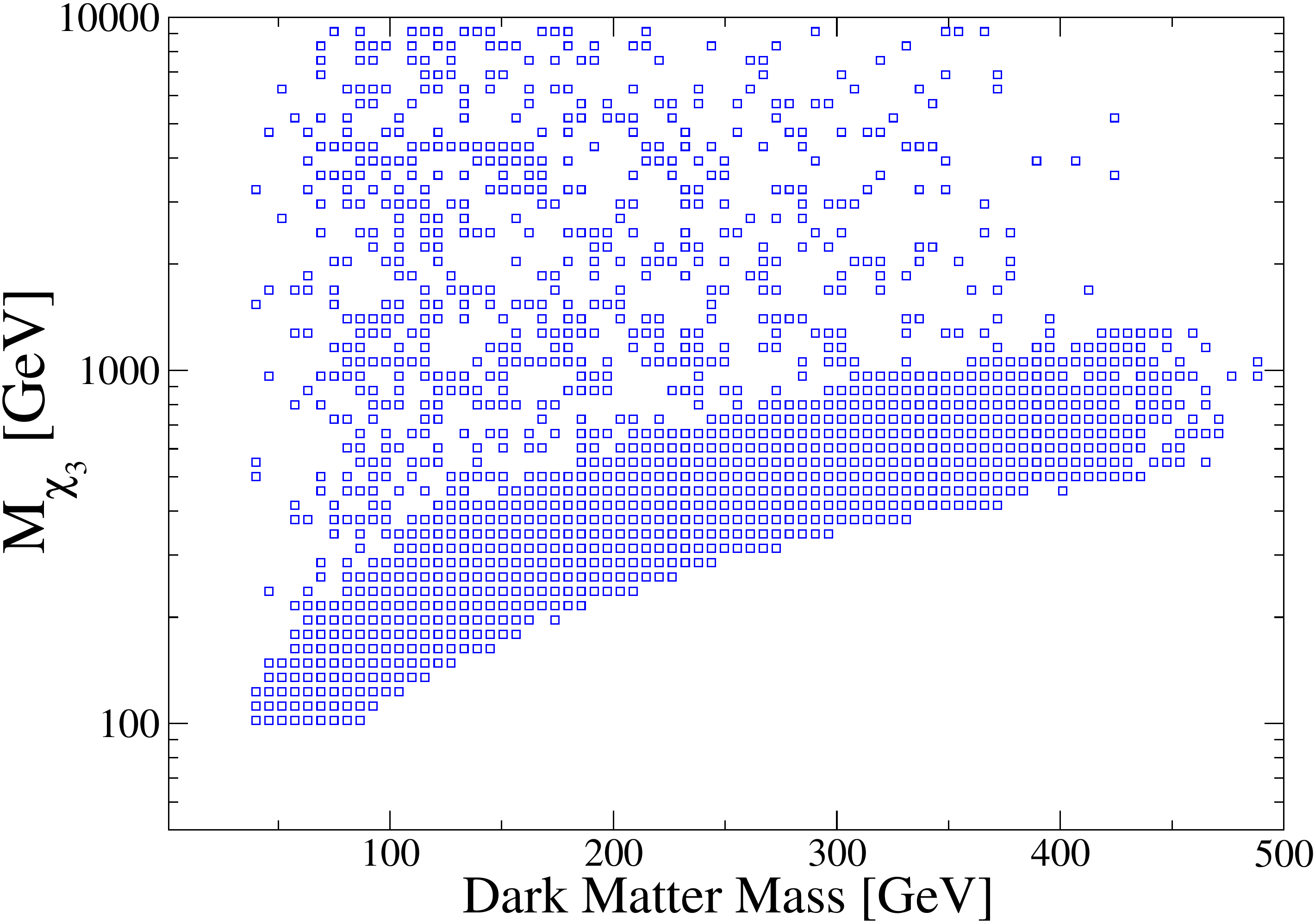}
  \end{tabular}
  \label{fig:masses}
  \caption{Left: The viable models projected onto the plane ($m_\phi$,$m_{\chi_1}$). The relic density constraint favors a $\chi_1$ not much heavier than the dark matter ($\phi$). Right: The viable models projected onto the plane ($m_\phi$,$m_{\chi_3}$). Notice that the heaviest odd fermion, $\chi_3$, rarely has a mass above 1 TeV. }
\end{figure}

The relic density depends strongly on the masses of the dark matter particle and of the Dirac fermion that mediates the annihilation processes.  The left panel of Fig.\ 5 displays the ratio $m_E/m_{\phi_1}$ as a function of the dark matter mass for our sample of viable points. The lower limit on $m_E/m_{\phi_1}$ is $1.2$ (to prevent coannihilations) and we see from the figure that it tends to that value at the highest dark matter masses. In any case, $m_E$ is never much larger than the dark matter mass. The dark matter constraint thus requires the existence of a charged fermion with a mass  below  $600$ GeV.

Two of the three neutral fermions have a mass very close to $m_E\sim \mu$ as a result of the small mixing induced by the parameters $\beta_{1,2}\ll 1$ (the mixing among the neutral fermions is very small). The other neutral fermion, with mass $\sim m_S$, can have a much larger mass.  The right panel of Fig.\ 5 displays the dark matter mass versus the mass of the heaviest neutral fermion, $M_{\chi_3}$. Even though this mass can reach its upper limit ($10$ TeV), in most models it tends to be close to  the dark matter mass. The spectrum in this setup is thus rather compressed.

\subsection{Implications for LFV processes}
\begin{figure}[tb]
  \centering
  \label{fig:taueg}
  \includegraphics[scale=0.45]{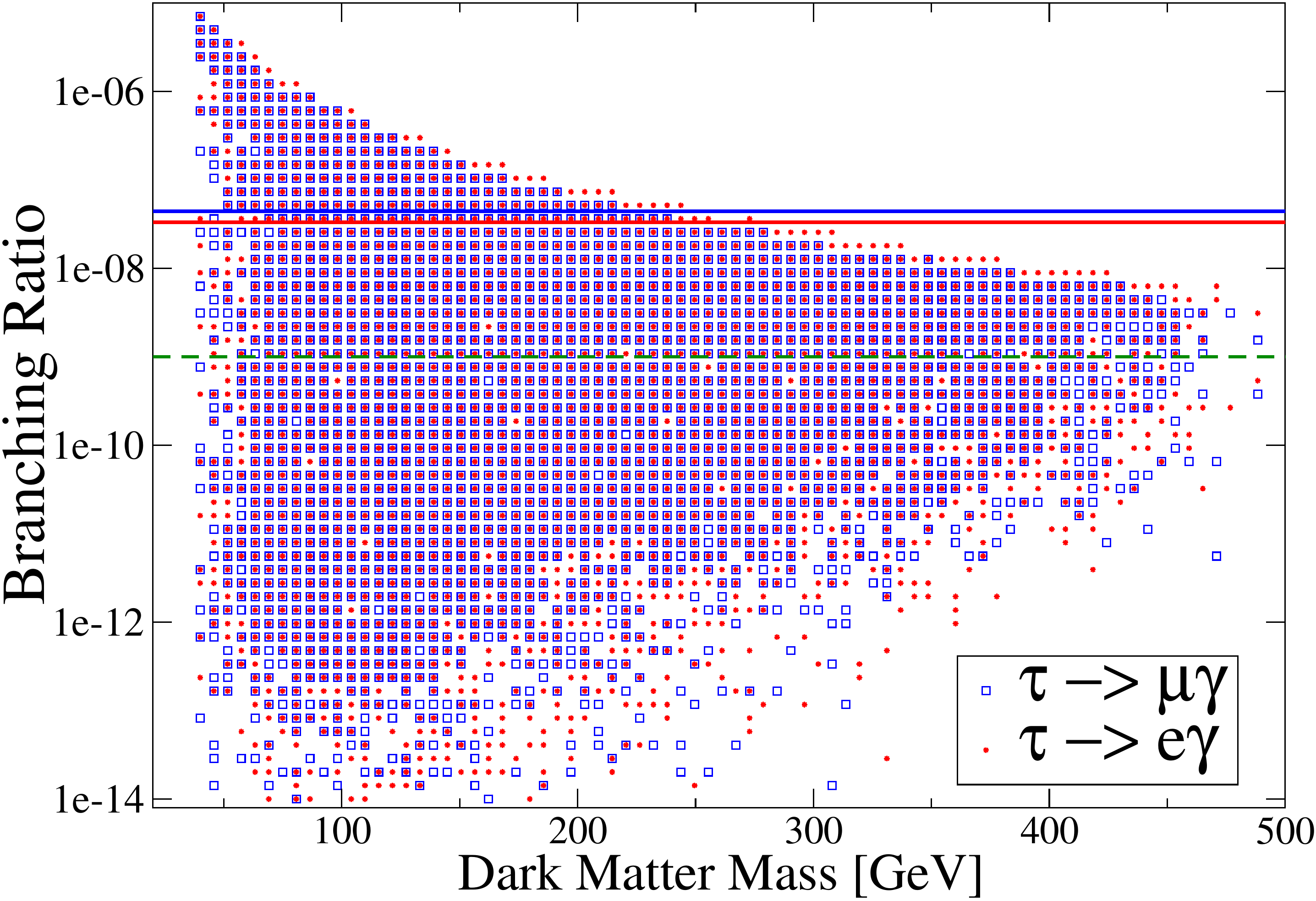}
  \caption{The branching ratios $\tau\to \mu\gamma$ (blue squares) and $\tau\to e \gamma$ (red stars) as a function of the dark matter mass for our sample of viable models. The solid lines show the current bound for each process while the dashed line corresponds to the expected future sensitivity.}
\end{figure}

In this section we study in  detail the predicted rates for the most relevant LFV processes and compare them against current limits and expected sensitivities in planned experiments. As we will see, this scenario predicts sizable rates for several LFV processes, particularly $\mu\to 3e$ and CR($\mu-e$).  These processes, in fact, provide the most promising way of probing this model in the near future.

Let us begin our analysis with LFV $\tau$ decays. Figure 6 displays the branching ratios for the decay processes $\tau\to\mu\gamma$ (blue squares) and $\tau\to e \gamma$ (red stars) as a function of the dark matter mass.  Both processes feature the same behaviour  -- its maximum value decreasing with the dark matter mass --  and similar values at a given mass. These branching ratios vary over a wide range at low masses (from $10^{-5}$ to $10^{-13}$ or so) but tend to concentrate between $10^{-8}$ and $10^{-10}$ at higher  dark matter masses. For comparison, current limits are also shown in the figure as solid lines,  as well as the expected future sensitivity for both processes (dashed line). Interestingly, we see that current limits can be violated for dark matter masses below $250$ GeV or so. In addition, future experiments  will be able to probe a significant fraction of models over the entire range of dark matter masses.

\begin{figure}[tb]
  \centering
  \label{fig:mue}
  \includegraphics[scale=0.45]{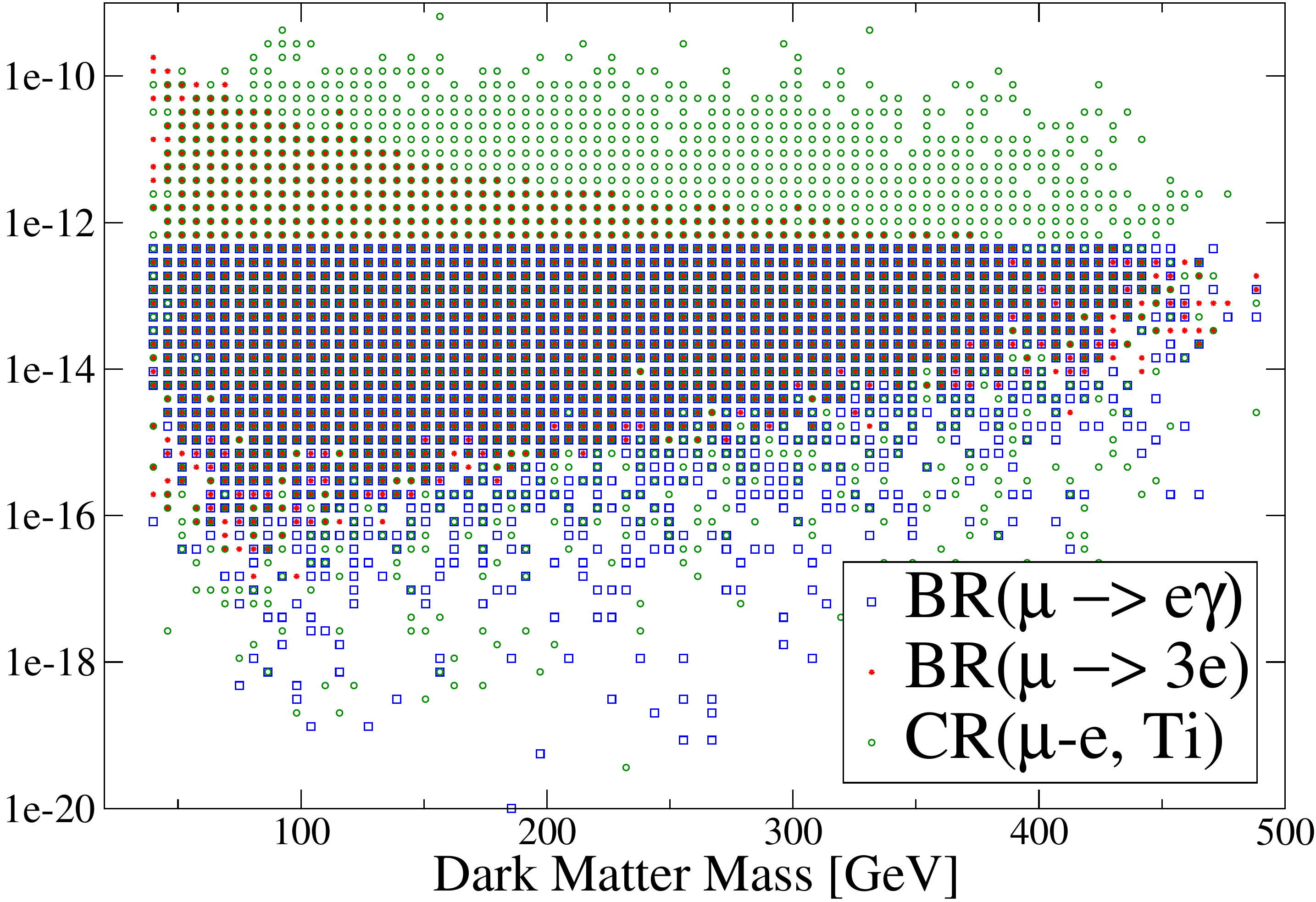}
  \caption{The rates of the most relevant $\mu$-$e$ LFV processes as a function of the dark matter mass  for our sample of viable models.}
\end{figure}

Let us now switch to $\mu$-$e$ LFV processes, which are typically more relevant. Figure 7 shows, as a function of the dark matter mass, the $\mueg$ (blue squares) and $\mu\to 3e$ (red stars) branching ratios as well as the $\mu$-$e$ conversion rate in Titanium (green circles). Notice that whereas the current limit on BR($\mu\to 3e$) can be violated only at low masses, the one on CR($\mu\mbox{-}e$, Ti) can be exceeded over the entire viable range of dark matter masses. It is also important to stress that the rates of all these $\mu\mbox{-}e$ LFV processes do not extend to arbitrarily low values, and tend instead to lie up to few orders of magnitude below present bounds.

\begin{figure}[tb]
  \begin{tabular}{cc}
  \includegraphics[scale=0.23]{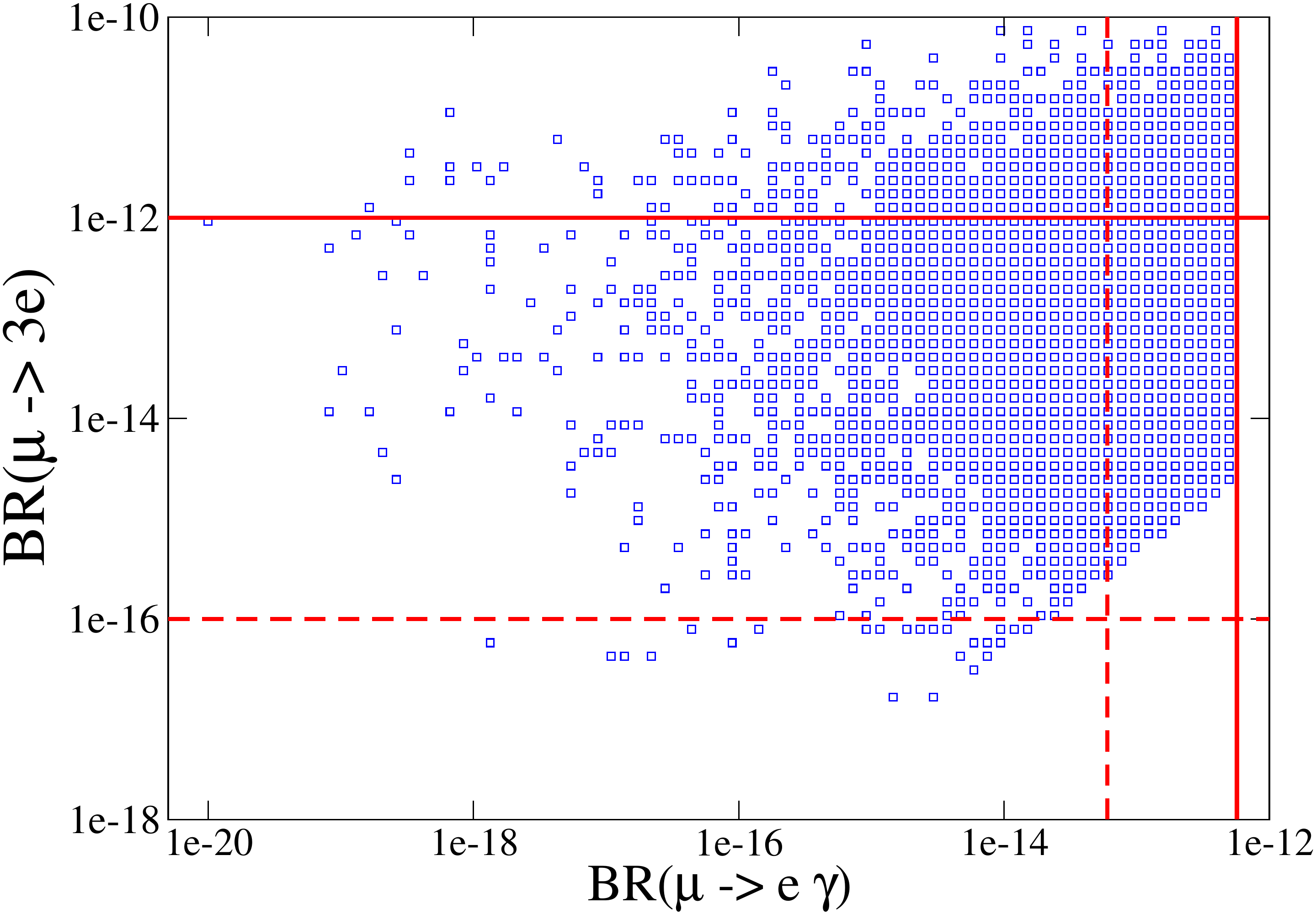} & \includegraphics[scale=0.23]{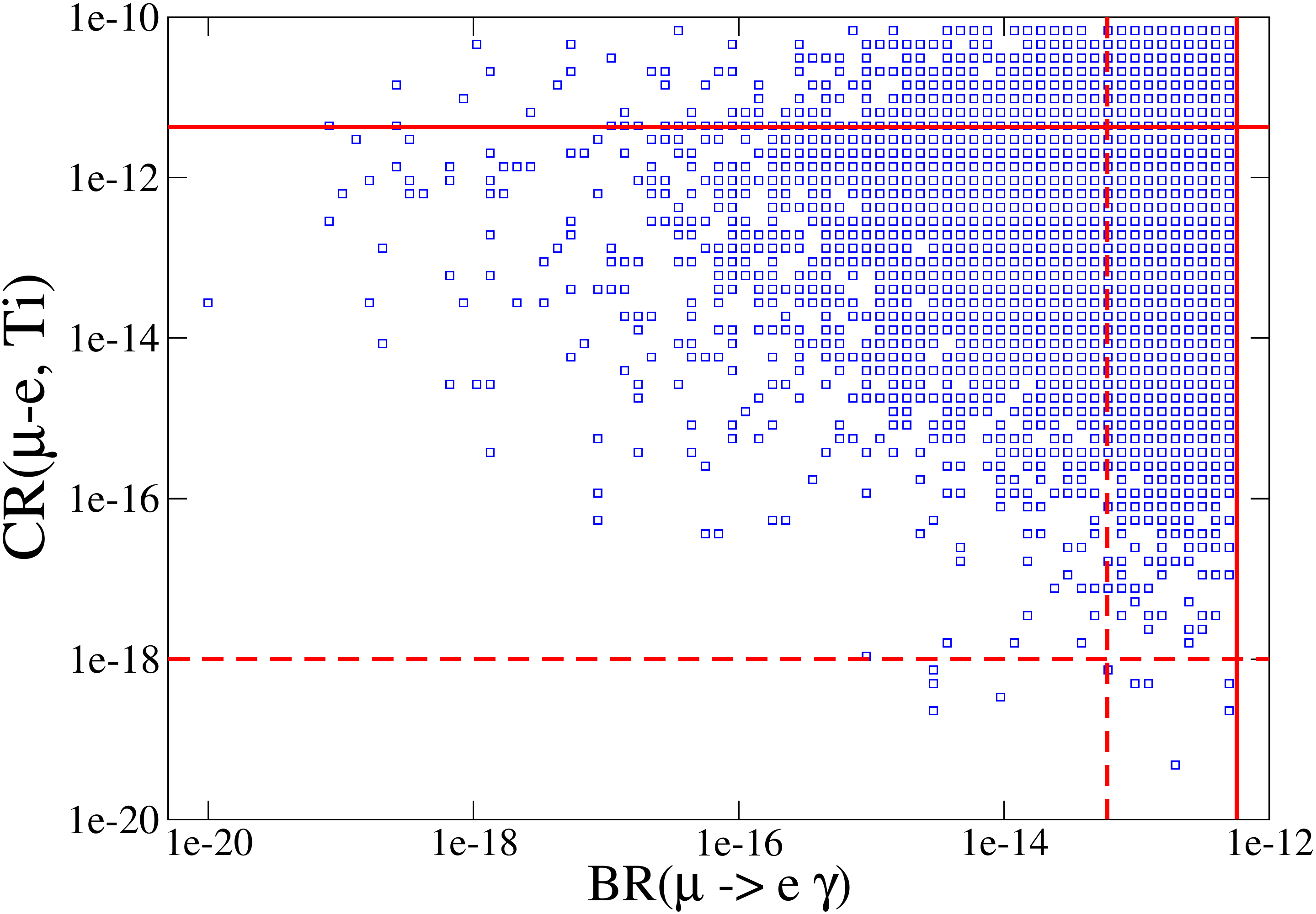}
  \end{tabular}
  \label{fig:muegmu3e}
  \caption{Left: The correlation between $\mu\to e\gamma$ and $\mu\to 3 e$ for our sample of viable models. Right: The correlation between $\mu\to e\gamma$ and $\mu$-$e$ conversion in Titanium. }
\end{figure}

Figure 8 explores the correlations between the different $\muec$ LFV processes in this model. The left panel shows the viable points in the plane BR($\mueg$) vs.\ BR($\mute$), whereas the right panel shows  BR($\mueg$) vs CR($\muec$, Ti). We also displayed, for each process, its current limit (solid lines) and its expected future sensitivity in planned experiments (dashed lines).  As we have already observed in the previous figure,  current limits on $\mu\to 3e$ and $\muec$ conversion in nuclei are not necessarily fulfilled and can be violated by more than two orders of magnitude. Thus, satisfying the $\mueg$ limit does not guarantee compatibility with other $\mu\mbox{-}e$ LFV experiments. In other words, contrary to the naive expectation, $\mu\to 3 e$ and $\muec$ in nuclei may give a more stringent constraint than $\mu\to e \gamma$ for some models. Regarding future prospects, we see from the figure that the improvement in $\mu\to e\gamma$ though important will not be decisive as many models feature much smaller branching ratios. Completely different is the situation for $\mu\to 3e$ and $\muec$ in nuclei. There, the improvements will be more significant, and their impact will be crucial for this scenario. In fact, very few models  lie beyond the expected sensitivity of future $\mu\to 3e$ or $\muec$ conversion experiments. 

\begin{figure}[tb]
  \centering
  \label{fig:mu3evscr}
  \includegraphics[scale=0.45]{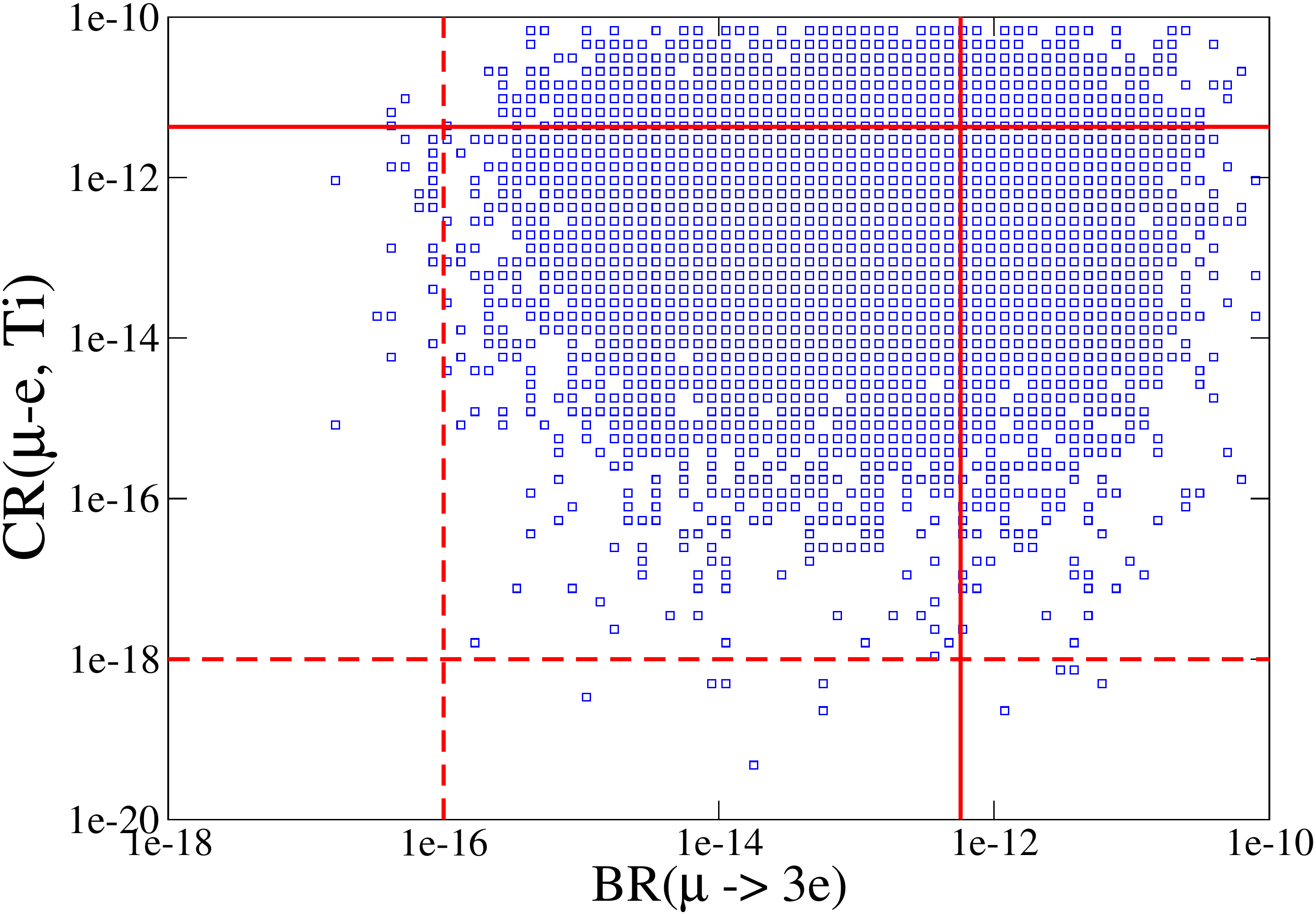}
  \caption{The correlation between $\mu\to 3e$ and $\mu$-$e$ conversion in nuclei  for our sample of viable models.  }
\end{figure}

To emphasize this point,  we plot instead BR($\mute$) versus CR($\muec$, Ti)  in Fig.\ 9. As noticed in the previous figure, only few points lie beyond the expected future sensitivity for either process. But now we can also observe that not a single point in our scan lies beyond the expected sensitivity for both processes. Future experiments searching for $\mu\to 3 e$ and $\muec$ conversion in nuclei therefore have the potential to probe most, if not all,  of the viable parameter space of this model.

\section{Discussion}
\label{sec:dis}
As we have seen, this model could be probed in future experiments via LFV processes. Other ways to test it include the direct \cite{Cushman:2013zza} and indirect detection \cite{Buckley:2013bha} of dark matter as well as searches at the LHC. In this section we give a brief look at these alternatives.

In our setup, the direct detection prospects are not very encouraging because the dark matter particle is leptophilic. Thus, it has a vanishing tree-level scattering cross section with nuclei. Radiative corrections will generate a non-zero cross section  but it will still be highly suppressed. At one-loop, both the spin-independent interaction (mediated by the Higgs) and the anapole-moment (mediated by the photon) can be generated in this model. Recently, these one-loop effects were calculated explicitly within a similar radiative model \cite{Ibarra:2016dlb} and it was found that  the predicted signal lies well below the expected sensitivity of future experiments for the anapole-moment and just barely within it for the spin-independent one.  In our scenario the situation is worse because  the spin-independent cross section  will be further suppressed by lepton masses or the $\beta_{1,2}$ parameters. We can safely conclude, therefore, that direct detection experiments cannot probe the viable parameter space of this model.

In spite of  the velocity-suppressed  annihilation cross section, the indirect detection of dark matter in this model is not entirely out of question. Notice that  the velocity suppression in this setup is stronger than for Majorana fermions, featuring a $v^4$ rather than a $v^2$ behavior. Given that $v\sim 10^{-3}$ for dark matter particles in the galactic halo, this $v^4$ suppression implies an annihilation rate today about twelve orders of magnitude below the thermal one ($\sigma v\sim 3\times 10^{-26}~\mathrm{cm^3s^{-1}}$. This suppression can however be avoided with the emission of an additional photon, the so-called internal bremsstrahlung process \cite{Bringmann:2007nk},  which gives rise to a gamma-ray feature at high energy. In \cite{Giacchino:2013bta,Toma:2013bka}, where this effect was studied for scalar dark matter,  it was found that,  depending on the specific parameters of the model, this feature could actually be observed in current and planned gamma-ray telescopes. The other indirect detection channels  -- neutrinos, positrons and antiprotons --  are not expected to play any role in constraining or testing this model.

As we have seen, our model predicts the existence of several particles with masses below the TeV scale. In particular, the charged fermion should have a  mass below $600$ GeV while the dark matter particle must be lighter than about $500$ GeV. Naively, one would think that such particles should be easily produced and detected at the LHC, but that is not really the case, as we now explain.

The LHC collaborations ATLAS and CMS have so far not performed any specific searches for models with radiative neutrino masses and dark matter candidates. Possible LHC constraints can therefore only be derived from analyses with leptonic signatures that are either model-independent or have been performed for models with similar additional physical states.

At first sight, our model bears some similarity with the Minimal Supersymmetric Standard Model (MSSM), where the heavy charged and neutral Z$_2$-odd fermions are the charginos $\tilde{\chi}_1^\pm$ and the (often mass-degenerate) neutralinos $\tilde{\chi}_2^0$. However, contrary to the MSSM our fermions do not decay via cascades to a ``golden'' charged three-lepton final state, a neutrino and two light neutralinos $\tilde{\chi}_1^0$ carrying away missing transverse energy $\not{\!\!E}_T$, but directly to two charged leptons, a charged lepton and a neutrino or two neutrinos plus two invisible scalar dark matter particles. The trilepton ATLAS \cite{Aad:2015eda} and CMS \cite{Khachatryan:2014qwa} analyses do therefore not apply.

Two-lepton final states with $\not{\!\!E}_T$ have also been analysed by ATLAS \cite{Aad:2014vma} and CMS \cite{CMS:2013bda}, but under specific assumptions for the masses of the intermediate sleptons. While these assumptions are absent in the dilepton searches for $Z'$ bosons \cite{ATLAS-CONF-2015-070,CMS:2015nhc}, the latter lack the $\not{\!\!E}_T$ criterion. Single-lepton final states have been analysed in searches for $W'$ bosons \cite{ATLAS-CONF-2015-063,CMS:2015kjy} which rely, however, on the partial reconstruction of an $s$-channel resonance from the observed transverse mass not applicable here.

The ATLAS and CMS analyses cited above thus currently do not impose any direct limits on our model. The LHC data would have to be reanalysed with full signal, background and detector simulations. And even if we were to do so, which is beyond the scope of the present paper, the LHC data are not expected to significantly constrain the parameter space of this model. 

\section{Conclusions}
\label{sec:conc}
We analyzed a TeV-scale extension of the SM that can simultaneously explain neutrino masses and the dark matter. The SM particle content was enlarged  with a vector-like doublet of SU(2), a singlet fermion, and two singlet scalars, all of them assumed to be odd under a Z$_2$ symmetry that guarantees the stability of the dark matter particle. In this scenario, neutrino masses are generated radiatively via one-loop processes mediated by the new fields. We examined the case where the dark matter particle  -- the lightest odd field --  is a scalar and its relic density is determined by its Yukawa interactions. In this case, interesting correlations appear between dark matter, neutrino masses, and lepton flavor violating processes. We studied analytically and numerically the phenomenology of this scenario. We found that the dark matter constraint can only be satisfied for a dark matter mass below $500$ GeV and a charged fermion mass below $600$ GeV.  We argued that neither existing collider searches at the LHC nor the dark matter direct or indirect detection searches can significantly constrain this scenario. A generic prediction of this setup is instead the existence of sizable rates for several LFV processes. In fact, they provide the most promising  way of probing this model in the near future. Future searches for $\mute$ and $\muec$ conversion in nuclei, in particular, have the potential to probe the entire parameter space of this model.

\section*{Acknowledgments}
The work of M.K.\ and D.L.\ is supported by the BMBF under contract 05H15PMCCA and by the Helmholtz Alliance for Astroparticle physics. C.Y.\ is supported by the Max Planck Society in the project MANITOP.
\bibliographystyle{apsrev}
\bibliography{darkmatter}
\end{document}